\documentclass[11pt,a4paper]{article}
\pdfoutput=1

\usepackage{jheppub}

\usepackage[english]{babel} % hyphenation and typographical rules for a language
\usepackage[utf8]{inputenc} % input encoding
\usepackage[T1]{fontenc} % fonts encoding
\usepackage{afterpage} % postpone a command
\usepackage{booktabs} % enhance the tables

\newenvironment{pagefigure}{\begin{figure}[!p]}{\afterpage\clearpage\end{figure}}
\renewcommand{\Re}{\mathop{\mathrm{Re}}}

\newcommand{\CNNS}{\text{C$\nu$NS}}
\newcommand{\Dmq}{\Delta m^2}
\newcommand{\Eps}{\epsilon}

\makeatletter
\DeclareRobustCommand\recite[1]{\begingroup\@fileswfalse\cite{#1}\endgroup}
\makeatother

\graphicspath{{figures/}}

\title{Neutrino Discovery Limit of Dark Matter Direct Detection
  experiments in the presence of Non-Standard Interactions}

\author[a,b,c]{M.~C.\ Gonzalez-Garcia,}
\author[d]{Michele Maltoni,}
\author[e,f]{Yuber F.\ Perez-Gonzalez}
\author[e]{and Renata Zukanovich Funchal}

\affiliation[a]{Departament de Física Quàntica i Astrofísica and
  Institut de Ciencies del Cosmos,\\
  Universitat de Barcelona, Diagonal 647, E-08028 Barcelona, Spain}

\affiliation[b]{Institució Catalana de Recerca i Estudis Avançats
  (ICREA),\\
  Pg. Lluis Companys 23, 08010 Barcelona, Spain.}

\affiliation[c]{C.N.\ Yang Institute for Theoretical Physics, Stony
  Brook University,\\
  Stony Brook, NY 11794-3840, USA}

\affiliation[d]{Instituto de Física Teórica UAM/CSIC, Calle de Nicolás
  Cabrera 13–15,\\
  Universidad Autónoma de Madrid, Cantoblanco, E-28049 Madrid, Spain}

\affiliation[e]{Departamento de Física Matemática, Instituto de
  Física, Universidade de São Paulo,\\
  R.\ do Matão 1371, CEP.\ 05508-090, São Paulo, Brazil}

\affiliation[f]{ICTP South American Institute for Fundamental Research
  \& Instituto de Física Teórica,\\
  Universidade Estadual Paulista, Rua Dr.\ Bento T.\ Ferraz 271,
  CEP.\ 01140-070, São Paulo, Brazil}

\emailAdd{maria.gonzalez-garcia@stonybrook.edu}
\emailAdd{michele.maltoni@csic.es}
\emailAdd{yfperezg@if.usp.br}
\emailAdd{zukanov@if.usp.br}

\abstract{The detection of coherent neutrino-nucleus scattering by the
  COHERENT collaboration has set on quantitative grounds the existence
  of an irreducible neutrino background in direct detection searches
  of Weakly Interacting Massive Dark Matter candidates.  This
  background leads to an ultimate discovery limit for these
  experiments: a minimum Dark Matter interaction cross section below
  which events produced by the coherent neutrino scattering will mimic
  the Dark Matter signal, the so-called \emph{neutrino floor}. In this
  work we study the modification of such neutrino floor induced by
  non-standard neutrino interactions within their presently allowed
  values by the global analysis of oscillation and COHERENT data.  By
  using the full likelihood information of such global analysis we
  consistently account for the correlated effects of non-standard
  neutrino interactions both in the neutrino propagation in matter and
  in its interaction in the detector.  We quantify their impact on the
  neutrino floor for five future experiments: DARWIN (Xe), ARGO (Ar),
  Super-CDMS HV (Ge and Si) and CRESST phase III (CaWO$_4$).
  Quantitatively, we find that non-standard neutrino interactions
  allowed at the $3\sigma$ level can result in an increase of the
  neutrino floor of up to a factor $\sim 5$ with respect to the
  Standard Model expectations and impact the expected sensitivities of
  the ARGO, CRESST phase III and DARWIN experiments.}


\preprint{YITP-SB-18-03, IFT-UAM/CSIC-18-025}

\arxivnumber{}

\begin{document}
\maketitle

%%%%%%%%%%%%%%%%%%%%%%%%%%%%%%%%%%%%%%%%%%%%%%%%%%%%%%%%%%%%%%%%%%%%%%%%%%%%%

\section{Introduction}

The existence of Dark Matter (DM) in the Universe is considered one of
the pillars of physics Beyond the Standard Model (BSM). However,
almost a century since it was suggested we remain ignorant of what it is
made of.  Nature has not been kind enough so far to provide us with an
unambiguous way to probe its character by means of its interaction
with a human made detector.  This is so despite the continuous efforts
of scientific and engineering ingenuity which has resulted into
experiments which are sensitive to weaker and weaker DM interactions
and to a larger range of the DM candidate masses, in particular, for
the well motivated DM models consisting of Weakly Interacting Massive
Particles (WIMPs)~\cite{Feng:2010gw}.

Neutrinos, on the other hand, have given us the first direct evidence
of BSM physics in the form of flavor transitions which were first
observed in natural neutrino fluxes produced in the Sun and in the
Earth atmosphere and have now been confirmed and precisely measured in
a variety of oscillation experiments~\cite{GonzalezGarcia:2007ib}.  As
usual in particle physics the signals of today are the backgrounds of
tomorrow, and the possibility of the coherent scattering of these
natural neutrinos with nucleus ($\CNNS$)~\cite{Freedman:1973yd} of the
target material giving a nuclear recoil, similar to what is expected
of a DM particle, will eventually constitute an irreducible background
for the WIMP discovery~\cite{Billard:2013qya} in the next generation
of DM direct detection experiments~\cite{Aprile:2015uzo,
  Angloher:2015eza, Aalbers:2016jon, Agnese:2016cpb, Aalseth:2017fik,
  Mount:2017qzi, Essig:2018tss}.  These experiments are expected to
have sensitivity to significantly lower WIMP-nucleon cross sections
thanks to a lower recoil energy threshold and/or a larger target mass.
This, in turn, will place these detectors closer to reach such
discovery limit also known as the \emph{neutrino floor}.  Thus, the
first observation of $\CNNS$ by the COHERENT
collaboration~\cite{Akimov:2017ade} with rates consistent with the SM
expectation bears important implications for them.

In this paper we want to quantify what is the range of variation of
the neutrino floor for such next generation experiments which is still
allowed by the present experimental constraints on the relevant
neutrino interactions.  One can phenomenologically study generic
contributions that affect $\CNNS$, and therefore can modify the
neutrino floor, by considering a model-independent approach where
flavor dependent four-fermion effective operators of neutrinos and
quarks are allowed.  These operators will induce neutral current (NC)
non-standard neutrino interactions (NSI) with the target nucleus at
the detector, and also modify in a non-trivial manner the matter
potential relevant to the flavor evolution of the solar and
atmospheric neutrinos.  For this reason not only COHERENT scattering
data but also results from oscillation experiments are relevant in the
determination of their allowed values.

There are several studies in the literature of how the expected rate
of neutrino background in direct DM detection experiments could be
altered by non-standard flavor-in\-de\-pen\-dent
contributions~\cite{Harnik:2012ni, Bertuzzo:2017tuf} as well as
flavor-dependent NSI's~\cite{Dutta:2017nht,
  AristizabalSierra:2017joc}.  Here we extend those works with a
detailed and statistically consistent evaluation of the impact of
NSI's on the neutrino floor at a variety of next generation DM
detection experiments. To do so we make use of the full likelihood
information of the global analysis of oscillation and COHERENT data to
consistently account for the correlated effects of the NSI's both for
neutrino propagation in matter and in its interaction in the detector.
In particular, we quantify their impact on the neutrino floor for five
future experiments: four with monoatomic detector targets (Xe used by
DARWIN, Ar by ARGO, Ge and Si by Super-CDMS HV) and one operating with
a molecular target (CaWO$_4$ for CRESST phase III).

This paper is organized as follows. In Sec.~\ref{sec:NSIL} we
introduce the relevant NC-NSI Lagrangian and summarize the results on
the constraints on the NC-NSI coefficients from the global analysis of
neutrino oscillation and COHERENT scattering data.  In
Sec.~\ref{sec:rate}, we use the full likelihood of such global
analysis to quantify the allowed modification to the differential
recoil rate of neutrino events at direct detection experiments by the
presence of these NC-NSI.  After describing briefly the full
statistical analysis used to obtain the DM discovery limit, in
Sec.~\ref{sec:NFexp} we present our results in terms of the range of
variation of the neutrino floor induced by the NC-NSI interactions at
the aforesaid future detectors using different target materials.
Finally, in Sec.~\ref{sec:conc} we draw our final conclusions.

%%%%%%%%%%%%%%%%%%%%%%%%%%%%%%%%%%%%%%%%%%%%%%%%%%%%%%%%%%%%%%%%%%%%%%%%%%%%%

\section{NC-NSI Lagrangian and constraints}
\label{sec:NSIL}

In this work we consider NC-NSI affecting neutral-current processes
relevant to neutrino coherent scattering in DM
experiments. Generically NC-NSI interactions can be parametrized in
the form:
\begin{equation}
  \label{eq:NSILagrangian}
  \mathcal L_\text{NSI} = -2 \sqrt{2} G_F \sum_{f,P,\alpha,\beta}
  \Eps_{\alpha\beta}^{f,P}(\bar\nu_\alpha\gamma^\mu P_L\nu_\beta)
  (\bar f\gamma_\mu P f) \,,
\end{equation}
where $G_F$ is the Fermi constant, $\alpha,\beta$ are flavor indices,
$P\equiv P_L, P_R$ and $f$ is a Standard Model (SM) charged
fermion. In this notation, $\Eps^{f,P}_{\alpha\beta}$ parametrizes the
strength of the new interaction with respect to the Fermi constant,
$\Eps_{\alpha\beta}^{f,P} \sim \mathcal{O}(G_X/G_F)$.

First we notice that if all possible operators in
Eq.~\eqref{eq:NSILagrangian} are added to the SM Lagrangian, the
Hamiltonian of the system which governs neutrino oscillations in
presence of matter is modified as
\begin{equation}
  \label{eq:Hnsi}
  H^\nu = H_\text{vac} + H_\text{mat}
  \equiv \frac{1}{2E_\nu} U_\text{vac}
  \begin{pmatrix}
    0 \\ &\Dmq_{21} \\ &&\Dmq_{31}
  \end{pmatrix}
  U_\text{vac}^\dagger
  + \sqrt{2} G_F N_e(x)
  \begin{pmatrix}
    1+\Eps_{ee} & \Eps_{e\mu} & \Eps_{e\tau} \\
    \Eps_{e\mu}^* & \Eps_{\mu\mu} & \Eps_{\mu\tau} \\
    \Eps_{e\tau}^* & \Eps_{\mu\tau}^* & \Eps_{\tau\tau}
  \end{pmatrix} \,,
\end{equation}
where $U_\text{vac}$ is the 3-lepton mixing matrix in
vacuum~\cite{Pontecorvo:1967fh, Maki:1962mu, Kobayashi:1973fv} and
$N_e(x)$ is the electron number density as a function of the distance
traveled by the neutrino in matter. For antineutrinos $H^{\bar\nu} =
(H_\text{vac} - H_\text{mat})^*$.  In Eq.~\eqref{eq:Hnsi} the
generalized matter potential depends on the \emph{effective} NC-NSI
parameters, $\Eps_{\alpha\beta}$, defined as
\begin{equation}
  \label{eq:eps-eff}
  \Eps_{\alpha\beta} = \sum_{f=u,d,e} Y_f (x) \, \Eps_{\alpha\beta}^{f,V} \,.
\end{equation}
Note that the sum only extends to those fermions present in the
background medium (up quarks, down quarks and electrons), and
$Y_f(x)=N_f(x)/N_e(x)$ is the average ratio of the number density of
the fermion $f$ to the number density of electrons along the neutrino
propagation path. In the Earth, the ratios $Y_f$ are constant to very
good approximation, while for solar neutrinos they depend on the
distance to the center of the Sun.  The presence of NC-NSI with
electrons, $f=e$, would affect not only neutrino propagation in
matter, as described in Eq.~\eqref{eq:Hnsi}, but also the
neutrino-electron cross section in experiments such as SK, Borexino
and reactor experiments; hence, for the sake of simplicity, in what
follows we consider only NC-NSI with either up quarks ($f=u$) or for
down quarks ($f=d$).  Also, it should be noted that only the
vector-like NC-NSI combination ($\Eps^V = \Eps^L + \Eps^R$) contribute
to the matter potential in neutrino oscillations. Those are precisely
the relevant couplings for $\CNNS$ in the approximation $E_R\ll
E_\nu$, where $E_R$ is the nuclear recoil energy and $E_\nu$ the
incoming neutrino energy.

In general, the matter potential in Eq.~\eqref{eq:Hnsi} contains a
total of 9 additional parameters per $f$: 3 diagonal real parameters
and 3 off-diagonal complex parameters (\textit{i.e.}, 3 additional
moduli and 3 complex phases). However, the evolution of the system
given by the Hamiltonian in Eq.~\eqref{eq:Hnsi} is invariant up to a
constant, so that oscillation experiments are only sensitive to the
differences between the diagonal terms in the matter potential.

As a consequence of the CPT symmetry, neutrino evolution is invariant
if the Hamiltonian in Eq.~\eqref{eq:Hnsi} is transformed as $H^\nu \to
-(H^\nu)^*$, see~\cite{GonzalezGarcia:2011my, Gonzalez-Garcia:2013usa}
for a discussion in the context of NC-NSI.  This transformation can be
realized by changing the oscillation parameters as
\begin{equation}
  \label{eq:osc-deg}
  \begin{aligned}
    \Dmq_{31} & \to -\Dmq_{31} + \Dmq_{21} = -\Dmq_{32} \,,
    \\
    \theta_{12} & \to \pi/2 - \theta_{12} \,,
    \\
    \delta_\text{CP} & \to \pi - \delta_\text{CP} \,,
  \end{aligned}
\end{equation}
and simultaneously transforming the NC-NSI parameters as
\begin{equation}
  \label{eq:NC-NSI-deg}
  \begin{aligned}
    (\Eps_{ee} - \Eps_{\mu\mu}) & \to -(\Eps_{ee} - \Eps_{\mu\mu}) - 2  \,,
    \\
    (\Eps_{\tau\tau} - \Eps_{\mu\mu}) &\to -(\Eps_{\tau\tau} - \Eps_{\mu\mu}) \,,
    \\
    \Eps_{\alpha\beta} &\to -\Eps_{\alpha\beta}^* \qquad (\alpha \neq \beta) \,,
  \end{aligned}
\end{equation}
see Refs.~\cite{Gonzalez-Garcia:2013usa, Bakhti:2014pva,
  Coloma:2016gei} for details.  In Eq.~\eqref{eq:osc-deg}
$\delta_\text{CP}$ is the leptonic Dirac CP phase, and we are using
here the parametrization conventions from
Ref.~\cite{Coloma:2016gei}. This degeneracy allows for a change in the
octant of $\theta_{12}$, which leads to the appearance of the
so-called LMA-D solution in the solar neutrino
analysis~\cite{Miranda:2004nb}, when combined with large diagonal
NC-NSI coefficients $\Eps_{ee} - \Eps_{\mu\mu} \sim - 2$, and it also
implies an ambiguity in the neutrino mass ordering.  Being an
intrinsic degeneracy of the full OSC+NSI Hamiltonian for neutrino
propagation it cannot be resolved by neutrino oscillation experiments
alone.

The presence of the NC-NSI's also affects the inelastic interaction of
neutrinos and therefore neutrino scattering experiments are sensitive
to some combinations of $\Eps_{\alpha\beta}^{f,P}$, depending on
whether the scattering takes place with nuclei or electrons, the
number of protons and neutrons in the target nuclei, and other
factors.  Thus altogether current experimental constraints on
vector-like NC-NSI parameters include those obtained from a global fit
to oscillation data~\cite{Gonzalez-Garcia:2013usa} as well as those
obtained from results from neutrino scattering data.
For oscillation constraints on NC-NSI parameters we refer to the
comprehensive global fit in the framework of $3\nu$ oscillation plus
NC-NSI with up and down quarks performed
in~\cite{Gonzalez-Garcia:2013usa} which we briefly summarize here for
completeness. All oscillation experiments except SNO are only
sensitive to vector NC-NSI via matter effects as described above.
There is some sensitivity of SNO to axial couplings in their NC data,
so for the sake of simplicity the analysis in
Ref.~\cite{Gonzalez-Garcia:2013usa} and all combinations that we will
present in what follows are made under the assumption of purely
vector-like NC-NSI.
The fit includes data sets from KamLAND reactor
experiment~\cite{Gando:2010aa} and solar neutrino data from
Chlorine~\cite{Cleveland:1998nv}, Gallex/GNO~\cite{Kaether:2010ag},
SAGE~\cite{Abdurashitov:2009tn}, Super-Kamiokande~\cite{Hosaka:2005um,
  Cravens:2008aa, Abe:2010hy, Smy:2012, Pik:2012qsy}
Borexino~\cite{Bellini:2011rx, Bellini:2008mr} and
SNO~\cite{Aharmim:2006kv, Aharmim:2005gt, Aharmim:2008kc,
  Aharmim:2011vm}, together with atmospheric neutrino results from
Super-Kamiokande phases 1--4~\cite{skatm1-4}, long-baseline results
from MINOS~\cite{Adamson:2013whj, Adamson:2013ue} and
T2K~\cite{Ikeda:2011}, and reactor results from
CHOOZ~\cite{Apollonio:1999ae}, Palo Verde~\cite{Piepke:2002ju}, Double
CHOOZ~\cite{Abe:2012tg}, Daya Bay~\cite{An:2013uza} and
RENO~\cite{Seo:2013}, together with reactor short baseline flux
determination from Bugey~\cite{Declais:1994ma, Declais:1994su},
ROVNO~\cite{Kuvshinnikov:1990ry, Afonin:1988gx},
Krasnoyarsk~\cite{Vidyakin:1987ue, Vidyakin:1994ut},
ILL~\cite{Kwon:1981ua}, G\"osgen~\cite{Zacek:1986cu}, and
SRP~\cite{Greenwood:1996pb}.

In principle, the analysis depends on the six $3\nu$ oscillations
parameters plus eight NC-NSI parameters per $f$ target. To keep the
fit manageable in Ref.~\cite{Gonzalez-Garcia:2013usa} only real NC-NSI
were considered and $\Dmq_{21}$ effects were neglected in the analysis
of atmospheric and long-baseline experiments. This renders the
analysis independent of the CP phase in the leptonic mixing
matrix. Furthermore in Ref.~\cite{Friedland:2004ah} it was shown that
strong cancellations in the oscillation of atmospheric neutrinos occur
when two eigenvalues of $H_\text{mat}$ are equal, and it is for such
case that the weakest constraints are placed.

For what concerns bounds from non-oscillation experiments, generically
NC-NSI parameters in gauge invariant models of new physics at high
energies are expected to be subject to tight constraints from charged
lepton observables ~\cite{Gavela:2008ra, Antusch:2008tz} and also
directly from collider data (see for example \cite{Davidson:2011kr}).
For those constructions the bounds on the NC-NSI of neutrinos are too strong
to lead to any observable effect in direct dark matter detection experiments. 
However, more recently it has been argued that viable gauge models with light
mediators (\textit{i.e.}, below the electro-weak scale) may lead to
observable effects in oscillations without entering in conflict with
other bounds~\cite{Farzan:2015doa, Farzan:2015hkd, Babu:2017olk} (see
also Ref.~\cite{Miranda:2015dra} for a discussion). Furthermore, for
very light mediators, bounds from high-energy neutrino scattering
experiments such as CHARM~\cite{Dorenbosch:1986tb} and
NuTeV~\cite{Zeller:2001hh} do not apply. 
Conversely, even for models with such very light mediators bounds on
the vector-like NC-NSI can be imposed by the direct observation of
coherent neutrino-nucleus scattering, \textit{i.e.}, the same reaction
which would give rise the neutrino floor in the DM direct detection
experiments. The modification to the $\CNNS$ cross section is given
by~\cite{Barranco:2005yy}
\begin{equation}
  \label{eq:recoil_NSI}
  \left. \frac{d\sigma^\nu(\nu_\alpha)}{dE_R} \right|_\text{NSI}
  = [\mathcal{Q}_\text{NSI}^\alpha]^2 \mathcal{F}^2(E_R)
  \frac{G_F^2 m_N}{4\pi} \left(1-\frac{m_N E_R}{2 E_\nu^2}
  \right) ,
\end{equation}
where $\mathcal{F}(E_R)$ is the Helm form factor~\cite{Lewin:1995rx},
$m_N$ is the nuclear mass, and $E_\nu$ and $E_R$ are the neutrino and
recoil energies, respectively. The factor
$\mathcal{Q}_\text{NSI}^\alpha$ depends on the
$\Eps_{\alpha\beta}^{f,V}$ parameters
\begin{multline}
  [\mathcal{Q}_\text{NSI}^\alpha]^2 = 4\Bigg\lbrace
  \left[ N \left( -\frac{1}{2} + \Eps_{\alpha\alpha}^{u,V}
    + 2\Eps_{\alpha\alpha}^{d,V} \right)
    + Z \left( \frac{1}{2}-2\sin^2\theta_W + 2\Eps_{\alpha\alpha}^{u,V}
    + \Eps_{\alpha\alpha}^{d,V} \right) \right]^2
  \\
  + \sum_{\beta\neq\alpha} \left[ N \left( \Eps_{\alpha\beta}^{u,V}
    + 2\Eps_{\alpha\beta}^{d,V} \right) +
    Z \left( 2\Eps_{\alpha\beta}^{u,V} + \Eps_{\alpha\beta}^{d,V} \right) \right]^2
  \Bigg\rbrace \,.
\end{multline}
In this respect, the most relevant results are those from the COHERENT
collaboration which recently reported the observation of coherent
neutrino-nucleus scattering at $6.7\sigma$~\cite{Akimov:2017ade} with
neutrinos produced from pion decay at rest on a 14.6~kg CsI[Na]
detector target.  The observed interaction rate is in good agreement
with the SM prediction and therefore can be used to constrain NC-NSI.

\begin{figure}\centering
  \includegraphics[width=0.95\textwidth]{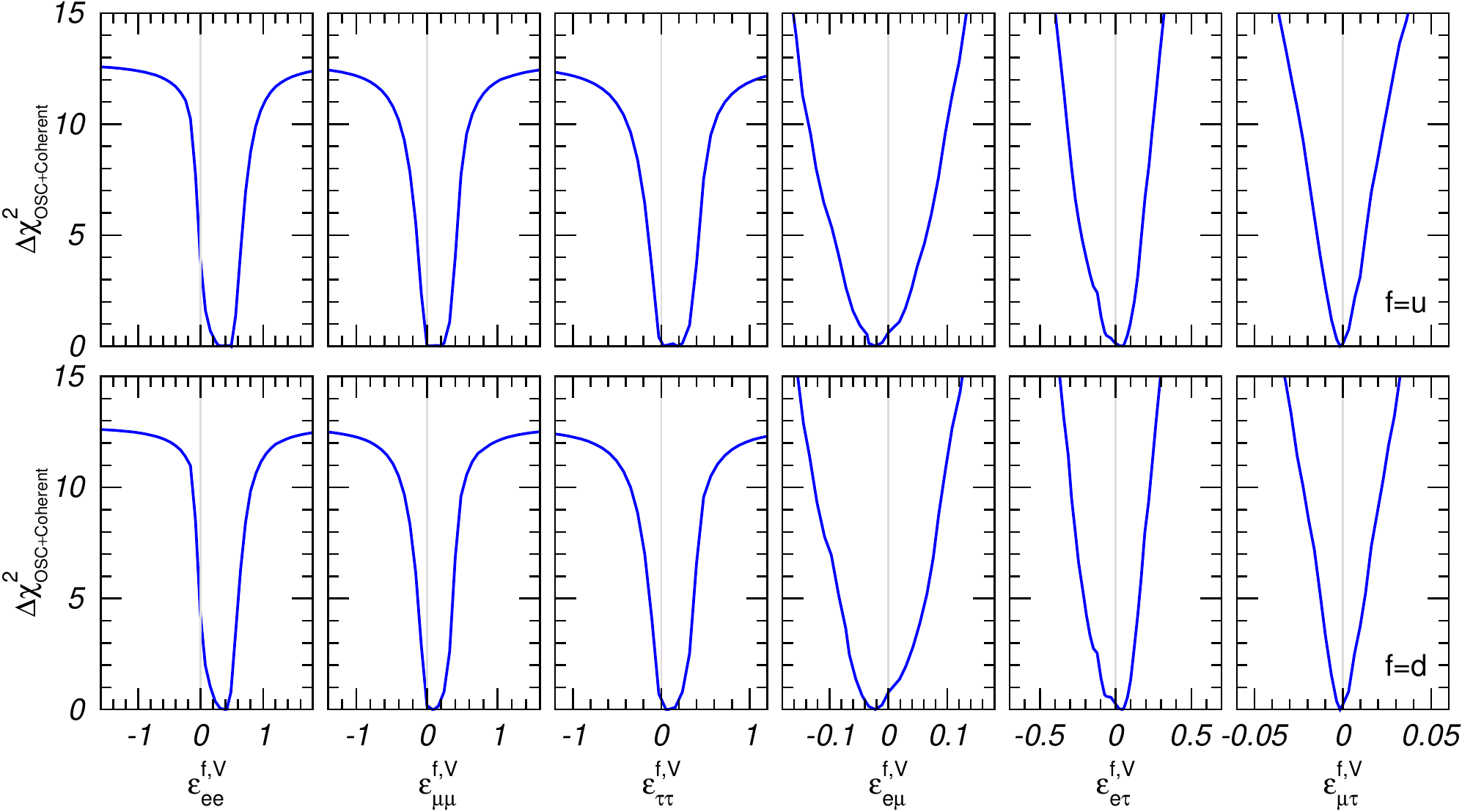}
  \caption{$\Delta\chi^2$ as a function of the NC-NSI parameters
    $\Eps_{\alpha\beta}^{f,V}$, for a global fit to oscillation
    experiments and COHERENT data (see text for details).}
  \label{fig:combo}
\end{figure}

In Ref.~\cite{Coloma:2017ncl} the COHERENT results were combined with
the neutrino oscillation constraints to produce model-independent
bounds on vector-like NC-NSI parameters which are relevant to the
evaluation of the neutrino floor for DM detection. In particular, it
was shown that after the inclusion of the COHERENT data the global
analysis excludes the intrinsic degeneracy in the NSI+OSC Hamiltonian
described above (\textit{i.e.}, the LMA-D solution) at $3.1\sigma$
($3.6\sigma$) CL for NC-NSI with up (down) quarks.  In addition, the
combination of oscillation and COHERENT allows to derive competitive
constraints on all NC-NSI parameters.  For illustration we show in
Fig.~\ref{fig:combo} the results from that combined fit by plotting
$\Delta\chi^2 \equiv \Delta\chi^2_\text{OSC+COH} =
\Delta(\chi^2_\text{OSC} + \chi^2_\text{COH})$ as a function of each
of the NC-NSI parameters $\Eps_{\alpha\beta}^{f,V}$, for $f=u$ (upper
panels) and $f=d$ (lower panels) after marginalization over the
undisplayed oscillation and NC-NSI parameters in each
panel.\footnote{Note that the global analysis slightly favors
  non-vanishing diagonal NC-NSI driven by the $2\sigma$ tension
  between the determination of $\Dmq_{21}$ from KamLAND and solar
  neutrino experiments (see, for example, Ref.~\cite{Esteban:2016qun}
  for the latest status on this issue).} As seen in the figure the
combination of oscillation and COHERENT results is able to impose
constraints on all the relevant vector-like NC-NSI coefficients even
when they are varied simultaneously in the analysis. In this respect,
it is important to notice that, in order to consistently obtain the
allowed effects on the neutrino floor at a given CL as presented in
the following section, we use the full multidimensional $\Delta\chi^2$
dependence on the NC-NSI and oscillation parameters so to correctly
account for the correlations among their allowed ranges.

%%%%%%%%%%%%%%%%%%%%%%%%%%%%%%%%%%%%%%%%%%%%%%%%%%%%%%%%%%%%%%%%%%%%%%%%%%%%%

\section{Neutrino rate at direct detection experiments}
\label{sec:rate}

Let us start by evaluating the impact of NC-NSI on the differential
recoil rate of solar and atmospheric neutrinos at direct detection
searches. In doing so one has to bear in mind that, unlike for SM
$\nu$ interactions~\cite{Billard:2013qya} and for the simplified BSM
models considered in Ref.~\cite{Bertuzzo:2017tuf}, NC-NSI are flavor
dependent and therefore in their presence one must take into
consideration the flavor of the neutrino fluxes arriving at the
detector.  This is incorporated by introducing the neutrino transition
probability between the source and the detector, so the differential
recoil rate takes the form:
\begin{equation}
  \label{eq:DRR-NSI}
  \left. \dfrac{dR}{dE_R} \right|_\nu =
  N_T \sum_{\nu_\alpha,\,\nu_\beta} \int_{E_\text{min}^\nu}
  \left. \dfrac{d\Phi}{dE_\nu} \right|_{\nu_\alpha} \,
  P(\nu_\alpha\to\nu_\beta, E_\nu) \,
  \left. \dfrac{d\sigma^\nu(\nu_\beta)}{dE_R} \right|_\text{NSI} dE_\nu,
\end{equation}
where $N_T$ corresponds to the total number of targets in the
detector, $d\Phi \big/ dE_\nu|_{\nu_\alpha}$ is the incoming flux of
neutrinos with flavor $\nu_\alpha$, $P(\nu_\alpha\to\nu_\beta, E_\nu)$
is the transition probability between the flavors
$\nu_\alpha\to\nu_\beta$, $d\sigma^\nu(\nu_\beta) \big/ dE_R
|_\text{NSI}$ is the flavor dependent $\CNNS$ cross section given in
Eq.~\eqref{eq:recoil_NSI} and $E_\text{min}^\nu$ is the minimum
neutrino energy that a neutrino must have to create a recoil with
energy $E_R$, that is
\begin{equation}
  E_\text{min}^\nu = \sqrt{\frac{m_N\,E_R}{2}} \,.
\end{equation}
For simplicity, we will only consider here the solar and atmospheric
contributions to the neutrino flux.  The Diffuse Supernova Neutrino
background~\cite{Beacom:2010kk}, expected but not yet observed, could
affect the discovery limit for very large
exposures~\cite{Ruppin:2014bra, OHare:2016pjy}. Other neutrino fluxes,
such as geoneutrinos~\cite{Billard:2013qya} and reactor
antineutrinos~\cite{Perez-Gonzalez:2017iso}, depend on the location of
the detector on the Earth, but in any case their intensity is very
small at practically all the important sites, thus making their
contribution negligible for the facilities and exposures analyzed
here.

The flavor transition probabilities for solar $\nu_e$ and atmospheric
$\nu_e$ and $\nu_\mu$ (and their antineutrinos) in the presence of
NC-NSI are obtained by solving the evolution equation:
\begin{equation}
  \label{eq:evol}
  i\frac{d}{dr}
  \begin{pmatrix}
    \nu_e\\
    \nu_\mu\\
    \nu_\tau
  \end{pmatrix}
  = H^\nu
  \begin{pmatrix}
    \nu_e\\
    \nu_\mu\\
    \nu_\tau
  \end{pmatrix},
\end{equation}
with $H^\nu$ given in Eq.~\eqref{eq:Hnsi}.

Solar neutrinos are electron neutrinos produced in the Sun core with
energy up to $\sim 20$~MeV. They are usually labeled as pp, pep, hep,
$^7$Be, $^8$B, $^{13}$N, $^{15}$O, and $^{17}$F neutrinos depending on
the reaction rate in which they are produced.  In what follows we will
consider the fluxes as predicted by the B16-GS98 Solar
Model~\cite{Vinyoles:2016djt}.  Its flavor transition probabilities
can be obtained in very good approximation in the one mass dominance
limit, $\Dmq_{31} \to \infty$ (which effectively means that
generically $G_F N_e \sum Y_f \Eps_{\alpha\beta}^{f,V} \ll \Dmq_{31} /
E_\nu$). In this approximation the survival probability $P_{ee}$ can
be written as~\cite{Kuo:1986sk, Guzzo:2000kx}
\begin{equation}
  \label{eq:peesun}
  P_{ee} = c_{13}^4 P_\text{eff} + s_{13}^4 \,,
\end{equation}
where $c_{ij} \equiv \cos\theta_{ij}$ and $s_{ij} \equiv
\sin\theta_{ij}$. The probability $P_\text{eff}$ can be calculated in
an effective $2\times 2$ model with the Hamiltonian $H_\text{eff} =
H_\text{vac}^\text{eff} + H_\text{mat}^\text{eff}$, with:
\begin{align}
  \label{eq:hvacsol}
  H_\text{vac}^\text{eff}
  &= \frac{\Dmq_{21}}{4 E_\nu}
  \begin{pmatrix}
    -\cos2\theta_{12} \, \hphantom{e^{-i\delta_\text{CP}}}
    & ~\sin2\theta_{12} \, e^{i\delta_\text{CP}}
    \\
    \hphantom{-}\sin2\theta_{12} \, e^{-i\delta_\text{CP}}
    & ~\cos2\theta_{12} \, \hphantom{e^{i\delta_\text{CP}}}
  \end{pmatrix} ,
  \\
  \label{eq:hmatsol}
  H_\text{mat}^\text{eff}
  &= \sqrt{2} G_F N_e(r)
  \begin{pmatrix}
    c_{13}^2 & 0 \\
    0 & 0
  \end{pmatrix}
  + \sqrt{2} G_F N_e(r)\sum_f Y_f(r)
  \begin{pmatrix}
    -\Eps_D^{f\hphantom{*}} & \Eps_N^f \\
    \hphantom{+} \Eps_N^{f*} & \Eps_D^f
  \end{pmatrix} .
\end{align}
where for $\delta_\text{CP}$ we have used the convention of
Ref.~\cite{Coloma:2016gei}. In this scheme, the coefficients
$\Eps_D^f$ and $\Eps_N^f$ are related to the original parameters
$\Eps_{\alpha\beta}^{f,V}$ by the following expressions:
\begin{align}
  \label{eq:eps_D}
  \begin{split}
    \Eps_D^f
    &= c_{13} s_{13}\, \Re\!\big( s_{23} \, \Eps_{e\mu}^{f,V}
    + c_{23} \, \Eps_{e\tau}^{f,V} \big)
    - \big( 1 + s_{13}^2 \big)\, c_{23} s_{23}\,
    \Re\!\big( \Eps_{\mu\tau}^{f,V} \big)
    \\
    & \hphantom{={}}
    -\frac{c_{13}^2}{2} \big( \Eps_{ee}^{f,V} - \Eps_{\mu\mu}^{f,V} \big)
    + \frac{s_{23}^2 - s_{13}^2 c_{23}^2}{2}
    \big( \Eps_{\tau\tau}^{f,V} - \Eps_{\mu\mu}^{f,V} \big) \,,
  \end{split}
  \\[2mm]
  \label{eq:eps_N}
  \Eps_N^f &=
  c_{13} \big( c_{23} \, \Eps_{e\mu}^{f,V} - s_{23} \, \Eps_{e\tau}^{f,V} \big)
  + s_{13} \left[
    s_{23}^2 \, \Eps_{\mu\tau}^{f,V} - c_{23}^2 \, \Eps_{\mu\tau}^{f,V*}
    + c_{23} s_{23} \big( \Eps_{\tau\tau}^{f,V} - \Eps_{\mu\mu}^{f,V} \big)
    \right].
\end{align}
The matter chemical composition of the Sun varies substantially along
the neutrino production region, with $Y_n$ dropping from about $1/2$
in the center to about $1/6$ at the border of the solar
core. Consequently the evolution is different when considering NC-NSI
with up or down quarks. Notice also that in the computation of the
transition probabilities one must account for the neutrino production
point distribution in the Sun which is determined for each neutrino
component (pp, pep, hep, $^7$Be, $^8$B, $^{13}$N, $^{15}$O, and
$^{17}$F) by the Solar Model.

Atmospheric neutrinos are electron and muon neutrinos and
antineutrinos produced by the interaction of cosmic rays with the
atmosphere and can extend to very large energies. In what follows we
consider the fluxes obtained by the FLUKA
collaboration~\cite{Battistoni:2005pd}. Only the low energy (sub-GeV)
part of their spectrum can be relevant as background in DM direct
detection experiments~\cite{Billard:2013qya}. Transition probabilities
for these low energy neutrinos present some unique
features~\cite{Peres:2003wd, Peres:2009xe} but generically Earth
matter effect are expected to have limited impact in their
evolution. Nevertheless, for the sake of consistency, we obtain the
flavor transition probabilities for the atmospheric neutrinos by
solving the $3\nu$ evolution equation (Eq.~\eqref{eq:evol}) using the
Earth matter density distribution described by the PREM
model~\cite{Dziewonski:1981xy}.  Since direct detection experiments do
not have directionality, one has to consider neutrino trajectories
arriving in the detector from any direction. Both the length of the
trajectory and the amount of Earth matter traversed will be different
for neutrinos arriving from different zenith angles, which makes their
flavor probability at the detector direction-dependent.  The
transition probability included in the differential recoil rate,
Eq.~\eqref{eq:DRR-NSI}, is then the angular average
\begin{equation}
  P^\text{atm}(\nu_\alpha\to\nu_\beta, E_\nu)
  = \frac{1}{2} \int_{-1}^1 d(\cos\eta)\,
  P^\text{atm}(\nu_\alpha\to\nu_\beta, E_\nu, \cos\eta).
\end{equation}

\begin{figure}\centering
  \includegraphics[width=\textwidth]{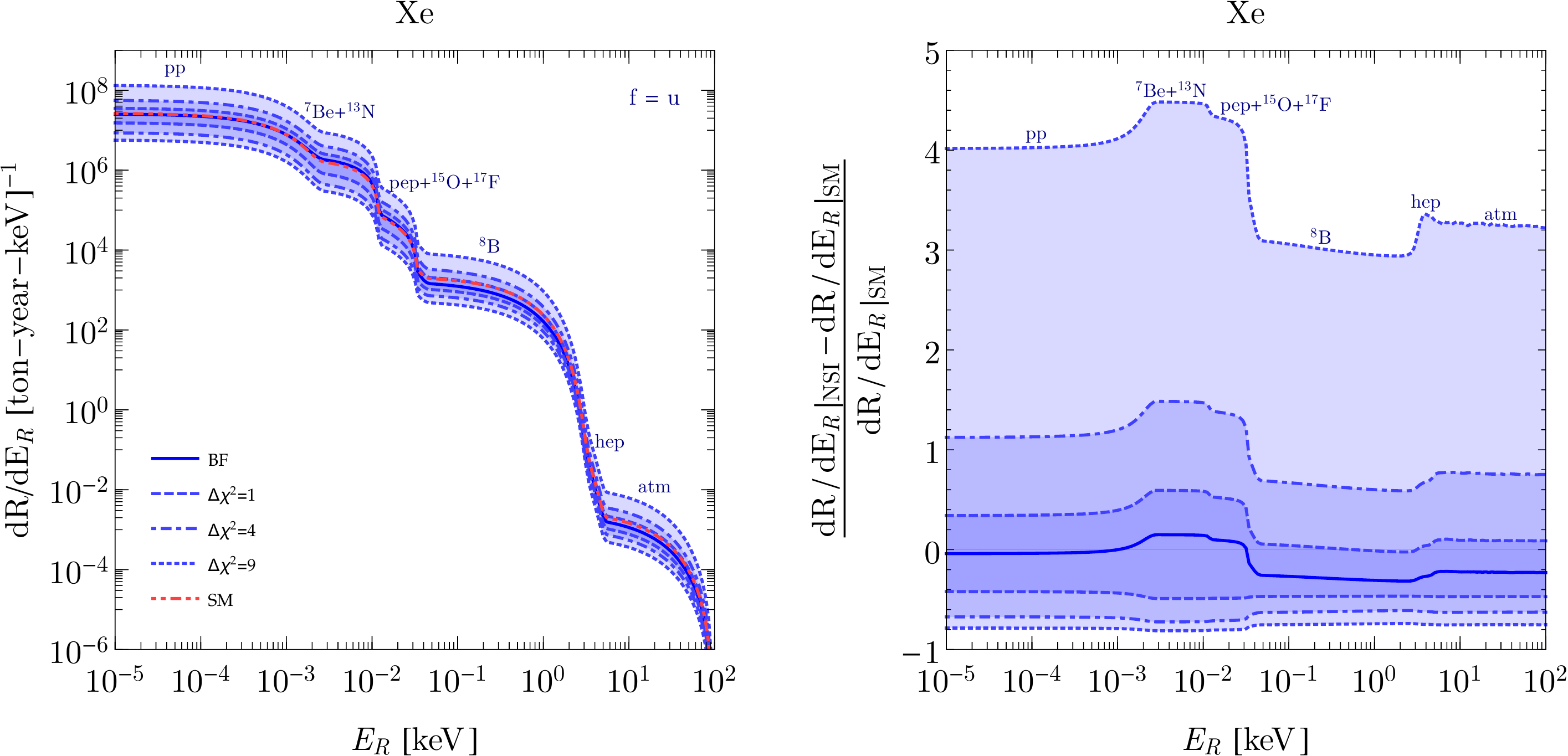}\\
  \vspace{2mm}
  \includegraphics[width=\textwidth]{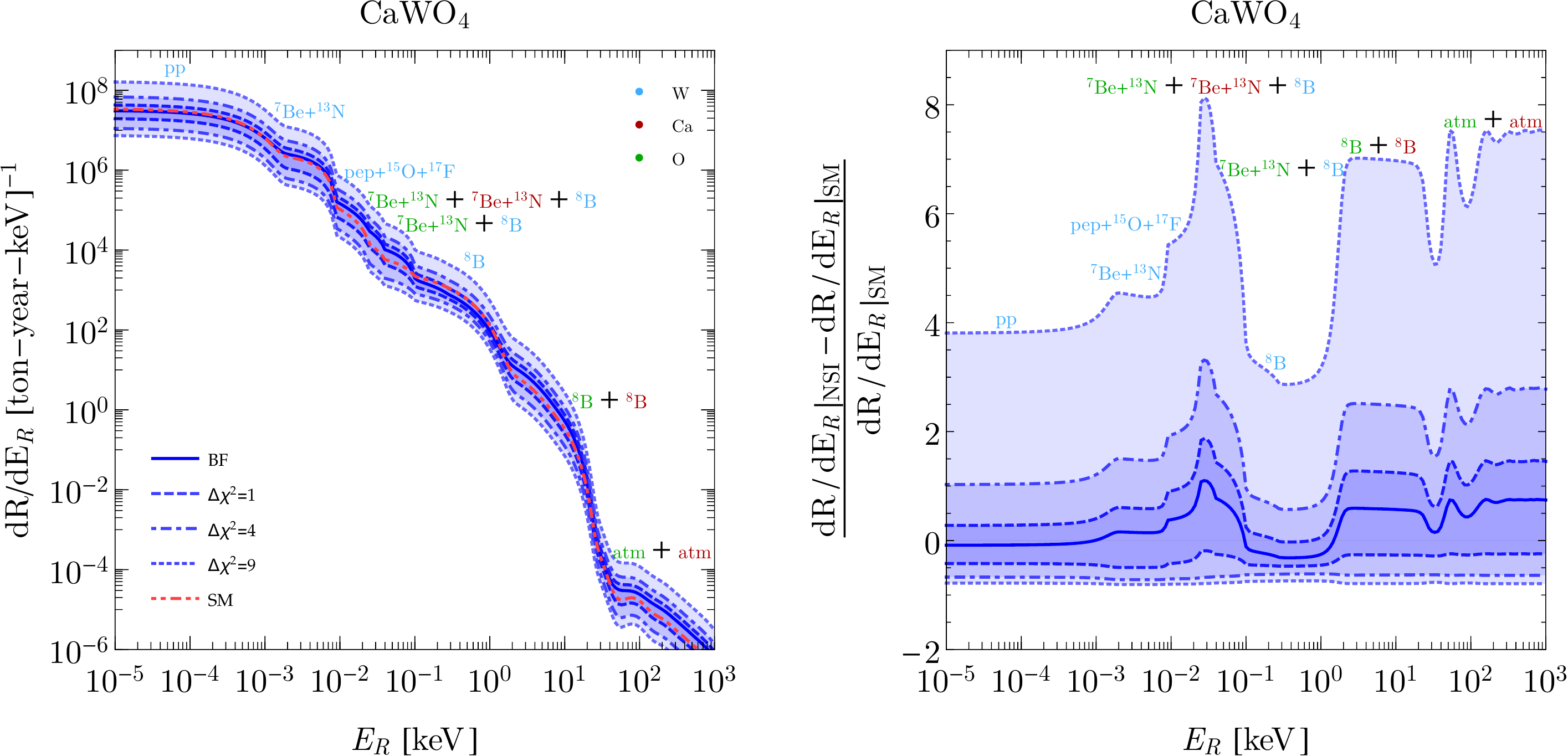}
  \caption{Expected differential recoil event rate (left) and the
    fraction of the NSI contribution compared to the SM one (right) as
    a function of the recoil energy $E_R$ for the solar and
    atmospheric neutrino fluxes.  In the upper (lower) panels the
    detector target is Xe (CaWO$_4$).  The full line corresponds to
    the prediction for the best fit NC-NSI and oscillation parameters
    of the analysis of oscillation+COHERENT data, and the shaded
    regions bordered by dashed, dot-dashed, and dotted lines
    correspond to predictions with NSI and oscillation parameters with
    $\Delta\chi^2_\text{OSC+COH} \leq 1$, 4, 9, respectively.}
  \label{fig:PdRXe}
\end{figure}

In Fig.~\ref{fig:PdRXe} we plot the differential recoil rate of solar
and atmospheric neutrinos as a function of the recoil energy for a
monoatomic target (Xe, upper panels) and for a molecular one
(CaWO$_4$, lower panels).  We only show the effects of NC-NSI for
$f=u$ as the cases for $f=d$ are very similar.  For illustration, we
label in the figure the dominant component of the neutrino flux
contributing at the different targets for each recoil energy range.
The shown ranges are obtained by scanning the multidimensional space
of NC-NSI and oscillation parameters allowed by the global analysis of
oscillation+COHERENT data at $1\sigma$, $2\sigma$, and $3\sigma$
($\Delta\chi^2_\text{OSC+COH} \leq 1$, 4, 9).  For each point in the
space of the allowed parameters we compute the predicted rate
consistently including the NC-NSI coefficients both in the oscillation
probabilities and in the interaction cross section.

We notice first that the prediction for best fit parameters is off the
SM expectation. As mentioned in Sec.~\ref{sec:NSIL} this is driven by
the $2\sigma$ tension between the determination of $\Dmq_{21}$ from
KamLAND and solar neutrino experiments.  From the figure we also see
that the allowed modification at $1\sigma$, $2\sigma$ and $3\sigma$
can result both in an increase or a reduction of the expected rate.
Quantitatively we find that with NC-NSI's allowed at $3\sigma$ by the
oscillation+COHERENT data the neutrino background rate can vary from a
factor $\sim \text{6-7}$ larger to a factor $\sim 0.2$ smaller than
what is predicted by the SM interactions, the larger enhancement
happening for a CaWO$_4$ detector.  Moreover, as seen in the figure,
the modification of the rate with respect to the SM prediction depends
on which nuclear reaction dominates the solar neutrino flux at the
given energy. This happens because of the NC-NSI in the matter
potential as well as the different distributions of neutrino
production points for the various nuclear reactions, which implies
that the modification of the transition probabilities depends on the
neutrino energy. However, altogether this is a relatively small
effect.  For solar neutrinos, the inclusion of the NC-NSI in the
matter potential leads to correction of up to $\sim 1\%$ ($\sim 12\%$)
in the rate of pp ($^8$B) neutrinos, while for atmospheric neutrinos
the difference induced by including NC-NSI in the propagation is of
order $\sim 1\%$. Finally, note that the peak structure that comes up
at $E_R \approx 50$~keV for CaWO$_4$ is a form factor
feature. Atmospheric neutrinos are relevant for recoil energies above
$\mathcal{O}(30~\text{keV})$.

%%%%%%%%%%%%%%%%%%%%%%%%%%%%%%%%%%%%%%%%%%%%%%%%%%%%%%%%%%%%%%%%%%%%%%%%%%%%%

\section{Results: NC-NSI effects on the direct detection discovery limit}
\label{sec:NFexp}

A WIMP direct detection experiment is based on a simple principle: if
a WIMP interacts with the quarks of a nucleus, it should be possible
to detect such interaction by measuring the nuclear recoil produced on
a target material. The negative results obtained so far have propelled
several proposals for future experiments~\cite{Aprile:2015uzo,
  Angloher:2015eza, Aalbers:2016jon, Agnese:2016cpb, Aalseth:2017fik,
  Mount:2017qzi}, which will have larger exposures and will be
sensitive to smaller WIMP masses. It is for such future experiments
that the solar and atmospheric neutrino background can represent a
limitation to sensitivity.

Bounds on the WIMP mass-cross section parameter space accessible by a
given experiment can be calculated using various statistical
methods. In this work we follow the approach of
Refs.~\cite{Billard:2011zj, Billard:2013qya, Ruppin:2014bra,
  OHare:2015mda, OHare:2016ows} which makes use of the likelihood
ratio test statistics to calculate the so-called \emph{discovery
  limit}, which is defined as the minimum cross section for a given
WIMP mass, $m_\chi$, for which a $3\sigma$ discovery is possible in
90\% of the hypothetical experiments.
In brief, we define a binned likelihood test statistics with $n_b =
50$ bins in the interval $[E_\text{th},\, 100~\text{keV}]$, where
$E_\text{th}$ is the energy threshold of the experiment:
\begin{equation}
  \label{eq:likelihood}
  \mathcal{L}(\sigma_{\chi n}, m_\chi; \phi_\nu)
  = \prod_{i=1}^{n_\text{b}} \mathcal{P}
  \bigg( N_\text{obs}^i \vert N^i_\chi
  + \sum_{j=1}^{n_\nu} N_\nu^i(\phi_\nu^j) \bigg)
  \times \prod_{j=1}^{n_\nu} \mathcal{L}(\phi_\nu^j) \, ,
\end{equation}
with $\mathcal{P}$ being the Poisson distribution function for
measuring a $N_\text{obs}^i$ number of events in bin $i$ from the
expected number of WIMP+neutrino events $N^i_\chi + \sum_{j=1}^{n_\nu}
N_\nu^i(\phi_\nu^j)$.  The sum over neutrino species $j$ includes all
solar and atmospheric components.
The expected number of neutrino events in bin $i$ from neutrino flux
component $\phi^j_\nu$ is trivially computed by integrating the
corresponding differential recoil rate in the bin,
\begin{equation}
  \label{eq:nu_events}
  \mathcal{N}_\nu^i(\phi_\nu^j) = \int_{E_i}^{E_{i+1}}
  \left. \frac{dR(\phi_\nu^j)}{dE_R} \right|_\nu \,
  \varepsilon(E_R)\, dE_R \,,
\end{equation}
while the expected number of WIMP events $N^i_\chi$ is
\begin{equation}
  \label{eq:wimp_events}
  \mathcal{N}_\chi^i = \int_{E_i}^{E_{i+1}}
  \left. \frac{dR}{dE_R}\right|_\chi \, \varepsilon(E_R)\, dE_R \,.
\end{equation}
The WIMP differential recoil rate is~\cite{Lewin:1995rx}
\begin{equation}
  \label{eq:DMrecoil}
  \left. \frac{dR}{dE_R} \right|_{\chi}
  = M \dfrac{\rho_0}{2\,\mu_n^2\,m_{\chi}}
  \sigma_{\chi n} (Z+N)^2\mathcal{F}(E_R)^2
  \int_{v_\text{min}} \frac{f(v)}{v}d^3v \,,
\end{equation}
where $\rho_0=0.3~\text{GeV} / c^2 / \text{cm}^3$ is the local DM
density and $\mu_n = m_n m_\chi / (m_n + m_\chi)$ is the reduced mass
of the WIMP-nucleon system~\cite{Agrawal:2010fh,
  Lewin:1995rx}. Furthermore, we assume a Maxwell-Boltzmann
distribution function for the WIMP velocity distribution in the
Earth's frame of reference $f(v)$, with the escape velocity
$v_\text{esc} = 544~\text{km/s}$, and the Earth's velocity
$v_\text{lab} = 232~\text{km/s}$~\cite{Lewin:1995rx}.  In
Eqs.~\eqref{eq:nu_events}--~\eqref{eq:wimp_events} $\varepsilon(E_R)$
is the detector efficiency function.

The terms $\mathcal{L}(\phi_\nu^j)$ in Eq.~\eqref{eq:likelihood}
correspond to prior factors included to parametrize the theoretical
uncertainties on the normalization of each neutrino flux component. We
take them to be Gaussian distributed with uncertainties $\sigma_j$
given by the B16-GS98 Solar Model (6\% for pp, 10\% for pep, 30\% hep,
6\% for $^7$Be, 12\% for $^8$B, 15\% for $^{13}$N, 17\% for $^{15}$O,
and 20\% for $^{17}$F), and the atmospheric flux calculations which we
conservatively take to be $20\%$~\cite{Honda:2011nf}.  These neutrino
flux normalizations will be treated here as nuisance parameters, while
the energy spectrum of each component is assumed to be
known.\footnote{This is a very good assumption for solar neutrinos as
  their energy dependence is determined by well-known phase space
  factors. For low energy atmospheric neutrinos the energy dependence
  uncertainty is also much smaller than the overall 20\% normalization
  considered here.} For the sake of simplicity we do not include in
our likelihood astrophysical uncertainties affecting the WIMP
signal~\cite{OHare:2016pjy}.

The test between the neutrino-only hypothesis, $H_0$, and the
neutrino+WIMP hypothesis, $H_1$, is performed by using the likelihood
ratio for a fixed WIMP mass
\begin{equation}
  \lambda(0)
  = \frac{\mathcal{L}(\sigma_{\chi n}=0, m_\chi; \mathring{\phi}_\nu)}
  {\mathcal{L}(\hat{\sigma}_{\chi n}, m_\chi; \hat{\phi}_\nu)} \,,
\end{equation}
where $\hat{\phi}_\nu$ and $\hat{\sigma}_{\chi n}$ are the neutrino
fluxes and WIMP-nucleon cross section values which maximize the
likelihood $\mathcal{L} (\hat{\sigma}_{\chi n}, m_\chi;
\hat{\phi}_\nu)$ while $\mathring{\phi}_\nu$ is obtained by maximizing
the likelihood function in the case of the null hypothesis,
$\mathcal{L}(\sigma_{\chi n}=0, m_\chi; \mathring{\phi}_\nu)$.
The significance of the WIMP signal (or, equivalently, of the
disagreement between $H_0$ and $H_1$) is $Z = \sqrt{-2 \ln
  \lambda(0)}$.  We construct its probability distribution function,
$p(Z\vert H_0)$, by MC simulating $2000$ experiments, creating for
each one a spectra for both WIMP and neutrinos and computing their
corresponding value of $Z$.  To determine the discovery limit we need
to find the smallest cross section for which 90\% of the experiments
have evidence larger than $3\sigma$, this is, the minimum value of
$\hat{\sigma}_{\chi n}$ for which
\begin{equation}
  \int_3^\infty p(Z\vert H_0) \, dZ  = 0.90 \,.
\end{equation}

\begin{figure}\centering
  \includegraphics[width=\textwidth]{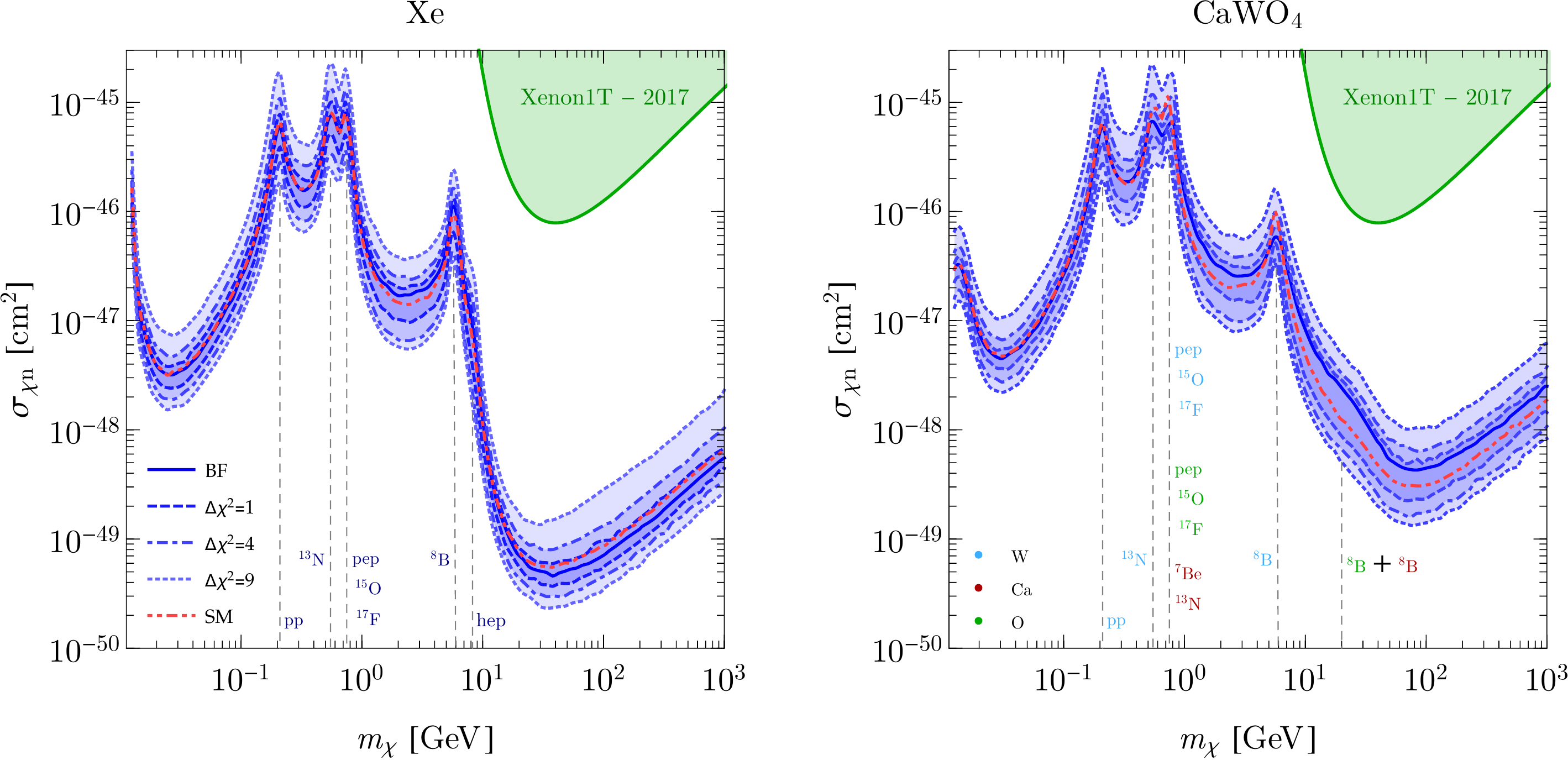}
  \caption{Impact on the discovery limit induced by NC-NSI couplings
    with up quarks for an \emph{idealized} detector with a Xe target
    (left) and a calcium tungstate (CaWO$_4$) target (right), a
    threshold of $E_\text{th} = 0.01$~eV, a $10^3$~ton-year exposure,
    and 100\% efficiency ($\varepsilon(E_R)=1$).  For illustration, we
    mark with vertical dashed lines the solar neutrino flux components
    dominating the background for some values of $m_\chi$ at each
    target; for the CaWO$_4$, the color indicates the contribution of
    the different elements.  The red dashed lines corresponds to SM
    predictions while the blue full lines to the best fit values of
    the parameters of the global analysis of oscillation+COHERENT
    data. The shaded regions bordered by dashed, dot-dashed, and
    dotted lines correspond to predictions with NC-NSI and oscillation
    parameters within $\Delta\chi^2_\text{OSC+COH}\leq 1$, 4, 9,
    respectively.  The green shaded region is excluded by the Xenon1T
    experiment at 90$\%$ CL~\protect\cite{Aprile:2017iyp}.}
  \label{fig:Generic_u}
\end{figure}

Clearly for an experiment with a given target, exposure and threshold
the presence of NC-NSI will affect the discovery limit as it modifies
the background neutrino event rate. To illustrate the maximal possible
size of the modification for a wide range of WIMP masses we start by
considering an idealized experiment with an extremely low threshold
($E_\text{th} = 0.01$~eV), a large exposure ($10^3$~ton-year) and a
100\% efficiency ($\varepsilon(E_R)=1$).  We plot the results in
Fig.~\ref{fig:Generic_u} for the same two targets considered in
Fig.~\ref{fig:PdRXe}, \textit{i.e.}, a monoatomic target of Xe and a
composed target made of calcium tungstate (CaWO$_4$). For the sake of
concreteness we consider NC-NSI with $f=u$ quarks but similar results
hold for NC-NSI with $f=d$.

As seen in Fig.~\ref{fig:Generic_u} the discovery limit curves exhibit
the well known peaks~\cite{Billard:2013qya, Ruppin:2014bra,
  OHare:2015mda, OHare:2016ows} corresponding to the region in the
parameter space which is significantly mimicked by the neutrino flux
component as labeled. For instance, the peak near a WIMP mass of 6~GeV
is related to the background created by $^8$B solar neutrinos.  Again,
the shown ranges are obtained by scanning the multidimensional space
of NC-NSI and oscillation parameters allowed by the global analysis of
oscillation+COHERENT data at $1\sigma$, $2\sigma$ and $3\sigma$
($\Delta\chi^2_\text{OSC+COH}\leq 1$, 4, 9).  For each point in the
allowed parameter space we compute the predicted neutrino background
rates in each bin, consistently including the NC-NSI coefficients both
in the oscillation probabilities and in the interaction cross section.
With such rates we obtain the corresponding discovery limit.  As
expected the inclusion of NC-NSI does not change the position of the
peaks since they do not modify the $E_\nu$ dependence of the $\CNNS$
cross section.  Also, as mentioned in the previous section the
neutrino floor predicted at the best fit point does not exactly
coincide with the SM prediction because of the preference of
oscillation data for a small but non-zero NC-NSI coefficients, which
helps reducing the tension between solar and KamLAND data in the
determination of $\Dmq_{21}$. As seen in the figure NSI-NC can lead to
an enhancement or a suppression of the discovery limit. Quantitatively
we find that within the $3\sigma$ allowed region the neutrino floor
can be modified between a factor $\sim 3$ ($^{13}$N neutrinos) to
$\sim 5$ (hep neutrinos) with respect to the expectation from the SM.
Finally, comparing the two panels in Fig.~\ref{fig:Generic_u} we see
that the main difference between the two targets appears in the region
of WIMP masses around $\sim 10$ GeV, and is a consequence of the
different recoil energy behavior of the neutrino event rates obtained
when overlapping the contribution from the different nuclear masses in
the composite target.

We now turn to estimate the possible impact of NC-NSI on the predicted
sensitivity of some proposed experiments. We will consider here four
facilities: three with monoatomic detector targets (DARWIN, ARGO and
Super-CDMS HV) and one using a molecular target (CRESST phase
III). Their targets, energy thresholds and exposures can be found in
Table~\ref{tab:DDExp}. The results are shown in Fig.~\ref{fig:Exp_u}
where we plot our estimate of the range of variation of the neutrino
floor induced by NC-NSI superimposed to the attainable bounds on the
WIMP-nucleon spin independent cross section as reported by each of the
experiments.  It is important to stress that this figure cannot be
used to make a final quantitative statement as the shown regions are
calculated under different assumptions. On the one hand we compute the
\emph{ultimate} neutrino floor, \textit{i.e.}, considering only the
neutrino irreducible background and for an experiment 100\% efficient
above its threshold.  On the other hand the attainable bounds quoted
by the experiments assume 30\% (DARWIN), 90\% (ARGO), 85\% (Super-CDMS
HV) and 70\% (CRESST phase III) efficiency above their energy
threshold. The curves also do not always correspond to the same
CL. Nevertheless, the comparison can be used to foresee which
experiments could potentially suffer a reduction of their sensitivity
reach because of the presence of NC-NSI.

\begin{table}\centering
  \catcode`?=\active\def?{\hphantom{0}}
  \begin{tabular}{ccccc}
    \toprule\toprule
    Target & Experiment & Threshold [keV] & Exposure [ton-year] & Reference
    \\
    \midrule
    CaWO$_4$ & CRESST -- Phase III
    & 0.1?? & 0.0027
    & \recite{Angloher:2015eza}
    \\
    Xe & DARWIN
    & 6.6?? & 200\hphantom{.00}
    & \recite{Aalbers:2016jon}
    \\
    Ge & Super-CDMS -- HV
    & 0.040 & 0.044?
    & \recite{Agnese:2016cpb}
    \\
    Si & Super-CDMS -- HV
    & 0.078 & 0.0096
    & \recite{Agnese:2016cpb}
    \\
    Ar & ARGO
    & 30\hphantom{.00} & 1000\hphantom{.0}
    & \recite{Aalseth:2017fik}
    \\
    \bottomrule
  \end{tabular}
  \caption{Proposed direct detection experiments considered in this work.}
  \label{tab:DDExp}
\end{table}

As can be seen from this figure, for DARWIN~\cite{Aalbers:2016jon}
(top-left panel) the SM discovery limit is below the expected
sensitivity, however after the inclusion of presently allowed NC-NSI
the neutrino floor can rise above the experimental sensitivity for
light WIMP masses $m_\chi \lesssim 20$~GeV.  This is also the case for
the ARGO~\cite{Aalseth:2017fik} (middle-left panel) and CRESST phase III 
\recite{Angloher:2015eza} (bottom panel) experiments for
which we find that NC-NSI can increase the neutrino floor above the
expected sensitivity reach for WIMP masses $m_\chi \gtrsim
50$~GeV and $m_\chi \sim 4 - 10$ GeV, respectively.  Thus if these experiments detect a signal 
compatible with a WIMP around those mass ranges, it could be also interpreted as a
neutrino event produced by flavor-dependent NC-NSI. Conversely the
results show that these experiments have the potential to place
meaningful bounds on NC-NSI.  Note, however, that in all the cases
presented here the floor has been somehow underestimated as a lower
(more realistic) efficiency would move the discovery limits up, hence
closer to the experimental reach.

\begin{pagefigure}\centering
  \includegraphics[width=0.8125\textwidth]{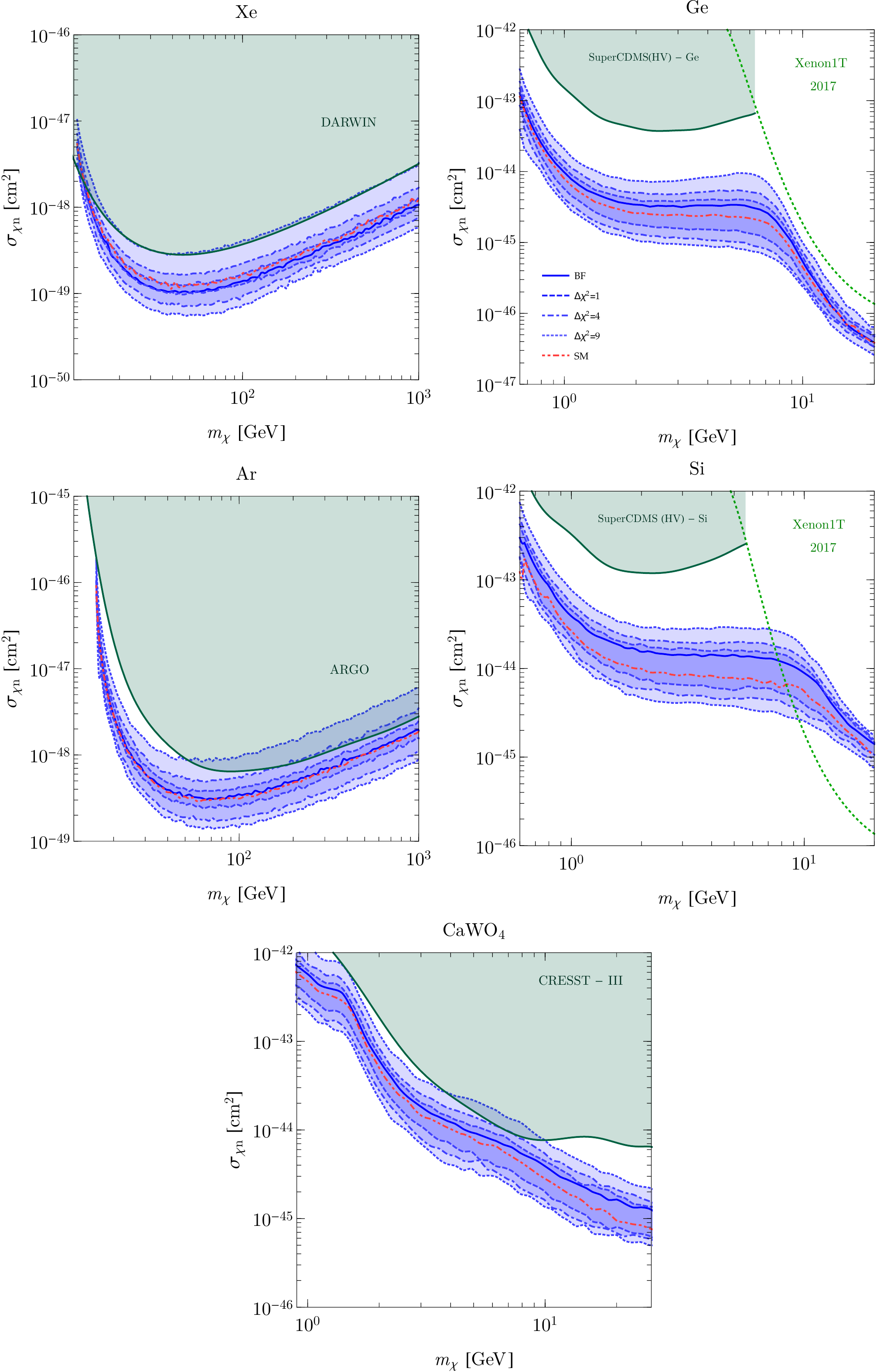}
  \caption{Impact of NC-NSI couplings with up quarks on the discovery
    limits of five different proposed experiments, DARWIN (top-left
    panel)~\recite{Aalbers:2016jon}, SuperCDMS - HV (Ge -- top-right,
    Si -- medium-right panels)~\recite{Agnese:2016cpb}, ARGO
    (medium-left panel)~\recite{Aalseth:2017fik} and CRESST phase III
    (bottom panel)~\recite{Angloher:2015eza}.  The red dashed line
    corresponds to the SM prediction while the blue full line to the
    best fit NSI and oscillation parameters of the analysis of
    oscillation+COHERENT data. The shaded regions bordered by dashed,
    dot-dashed, and dotted lines correspond to predictions with NSI
    and oscillation parameters within $\Delta\chi^2_\text{OSC+COH}
    \leq 1$, 4, 9, respectively.  In each panel we show in green, for
    comparison, the sensitivity region as quoted by each of these
    experiments in the given references. For reference we also show by
    a green dashed line the region excluded by Xenon1T at 90$\%$
    CL~\protect\cite{Aprile:2017iyp}.}
  \label{fig:Exp_u}
\end{pagefigure}

%%%%%%%%%%%%%%%%%%%%%%%%%%%%%%%%%%%%%%%%%%%%%%%%%%%%%%%%%%%%%%%%%%%%%%%%%%%%%

\section{Conclusions}
\label{sec:conc}

Next generation direct detection DM experiments are expected to
substantially increase the sensitivity to WIMP-nucleon cross sections
for a wide range of DM masses.  This will place them closer to the
region where solar and atmospheric neutrinos undergoing $\CNNS$ will
become an irreducible background.

While the first measurement of $\CNNS$ by the COHERENT experiment
observed scattering rates that seem to be in agreement with the SM
predictions, BSM contributions of NC-NSI with the target nucleus, at a
level that could impact future DM detectors, cannot be excluded. This
is true even if neutrino oscillation constraints are taken into
account.

We investigated in this work to what extent the future experimental
discovery limits of DM direct detection facilities could be
compromised by this flavor-dependent BSM. We used the full likelihood
information of the global analysis of neutrino oscillation experiments
and COHERENT scattering data to consistently account for the
correlated effects of NC-NSI in the neutrino propagation in matter as
well as in its interaction with the detector.

To be specific, we estimated the maximum effect which these still
allowed BSM interactions could have in experiments with different
targets (Xe in DARWIN, Ar in ARGO, Ge and Si in Super-CDMS HV, and
CaWO$_4$ in CRESST phase III).  We showed that the neutrino background
contribution can be up to about five times larger than what is
expected by the SM, certainly having impact on the ultimate reach of
the ARGO, CRESST phase III and DARWIN experiments.  In turn, these experiments may be
able to constraint NC-NSI to yet lower levels than currently allowed
by data.

%%%%%%%%%%%%%%%%%%%%%%%%%%%%%%%%%%%%%%%%%%%%%%%%%%%%%%%%%%%%%%%%%%%%%%%%%%%%%

\acknowledgments

This work was supported by Fundação de Amparo à Pesquisa do Estado de
São Paulo (FAPESP), by Conselho Nacional de Ciência e Tecnologia
(CNPq), by USA-NSF grant PHY-1620628, by EU Networks FP10 ITN ELUSIVES
(H2020-MSCA-ITN-2015-674896) and INVISIBLES-PLUS
(H2020-MSCA-RISE-2015-690575), by MINECO grants FPA2016-76005-C2-1-P,
FPA2012-31880 and MINECO/FEDER-UE grants FPA2015-65929-P and
FPA2016-78645-P, by Maria de Maetzu program grant MDM-2014-0367 of
ICCUB, by the ``Severo Ochoa'' program grant SEV-2016-0597 of IFT.

\bibliography{references}{}
\bibliographystyle{JHEP}

\end{document}